\documentclass [11pt, one column, a4paper] {article}

\usepackage[dvips]{epsfig}
\usepackage{amsmath}
\usepackage{amssymb}
\usepackage{graphicx}
\usepackage{bm}
\usepackage{epstopdf}
\usepackage[unicode=true,
 bookmarks=true,bookmarksnumbered=false,bookmarksopen=false,
 breaklinks=false,pdfborder={0 0 1},backref=false,colorlinks=false]
 {hyperref}
\usepackage{color}

\title{Features of planar Lee-Wick electrodynamics}

\author{L.H.C. Borges$^{1}$\thanks{email: luizhenriqueunifei@yahoo.com.br}, F.A. Barone$^{2}$\thanks{email: fbarone@unifei.edu.br}\\
{\small $^{1}$Universidade Federal do ABC, Centro de Ci\^encias Naturais e Humanas,}\\
{\small Rua Santa Ad\'elia, 166, 09210-170, Santo Andr\'e, SP, Brazil}\\
{\small $^{2}$IFQ - Universidade Federal de Itajub\'a, Av. BPS 1303, Pinheirinho,}\\
{\small Caixa Postal 50, 37500-903, Itajub\'a, MG, Brazil}}

\date {}

\begin {document}

\baselineskip=12pt

\maketitle

\begin{abstract}
In this letter we study some aspects of the planar Lee-Wick electrodynamics near a perfectly conducting line (unidimensional mirror). Specifically, the modified Lee-wick propagator due to the presence of a conducting line is calculated, and the interaction between the mirror and the point-like charge is investigated. It is shown that the behavior of this interaction is very different from the one already known for the $(3+1)$-dimensional Lee-Wick electrodynamics, where we have a planar mirror. It is also shown that the image method is not valid in planar Lee-Wick electrodynamics and the dimensional reduction yields a stronger taming of divergences.
\end{abstract}

Field theories with higher order derivatives have been intensively investigated in the literature. One of the main reasons to consider these kind of models is to improve renormalization properties and to remove ultraviolet divergences. In this scenario, the best-known and simplest gauge theory of this type it the one proposed by B. Podolsky and T. Lee and G. Wick \cite{PLW1,PLW2,PLW3,PLW4,PLW5}.

Some aspects of field theories with higher order derivatives have been investigated, for example, in issues which concern the point-charge self-energy \cite{CSE1,CSE2,CSE3,CSE4}, gravity theories \cite{g1,g2,g3,g4}, interactions between external sources \cite{DiracString4}, issues related to the Lee-Wick Standard Model \cite{sm1,sm2,sm3,sm4}, Pauli-Villars regularization \cite{tibes}, the presence of boundary conditions \cite{LeeWick,LHCFABEVE,casimirLW,Mbalz}, radiative corrections \cite{LHCFABCA}, among many others. We also highlight the relevance of field models with higher order derivatives in condensed matter physics, in the study of critical points \cite{PRL35} and phase transitions \cite{PRD86Ghosh}.

Planar physics, which describes field theories in $3d$ dimensions, is a subject of permanent interest due to their value in condensed matter systems. In this scenario, some very interesting physical properties can emerge in odd spacetimes. In the context of higher order derivatives involving scalar fields, we can mention, for instance, the study of quantum critical points with effective field theories with higher order derivatives \cite{AnnPhys310} and the role of field theories with higher order derivatives in the entanglement phase transitions \cite{SCIREP}. Furthermore, the planar QED is a very useful tool to describe the quantum Hall Effect \cite{PLB352,PRL53,PRB79,PRD48}, the graphene physics \cite{PTP128}, the behavior of (high temperature) cuprate superconductors \cite{PRB66,PRL86} and ultracold atoms \cite{PRL110}. Besides, the planar QED is used as a valuable toy model for the QCD \cite{PRL91} and the confinement phenomenon \cite{PRB77}.

The planar Lee-Wick quantum electrodynamics have been considered in a recent paper, where its quantization is performed in details in the Heisenberg picture \cite{Pimentel}. In this same context, the M\o ller scattering was studied in \cite{Montenegro}. In addition, the boson-boson interaction mediated by the planar Lee-Wick electrodynamics was considered in \cite{Accioly}. 

Therefore, it would very be welcome to have a better insight on the role of boundary conditions (mainly the ones related to material boundaries) imposed in the gauge field in the planar Lee-Wick electrodynamics, as well as the role of field sources in this theory.  

Investigations which concern the Lee-Wick electrodynamics in the vicinity of a perfectly conducting plate (two-dimensional mirror) were performed in $3+1$ dimensions in the work \cite{LeeWick}. More specifically, in this reference it was analysed the interaction between a stationary point-like charge and the conductor. In special, it was showed that the image method is not valid in $4d$ Lee-Wick electrodynamics. However, investigations of this type have not yet been carried out in the context of planar Lee-Wick electrodynamics. This topic is an interesting subject since planar models often show different properties in comparison with the correspondent ones obtained in $3+1$ dimensions. 

In this letter, we investigate some aspects of planar Lee-Wick electrodynamics near a perfect mirror, which in $2+1$ dimensions is just a perfectly conducting line. We compute the propagator for the Lee-Wick gauge field in the presence of a conducting line and analyse the interaction between a static point-like field source and the conducting line. We show that the behavior of this interaction is very different from the one already known in $3+1$ dimensions with a planar mirror. We also verify that image method remains not valid in $3d$ Lee-Wick electrodynamics. We show that the dimensional reduction yields a stronger taming of divergences in the Lee-Wick electrodynamics. Along the letter we will be working in a $(2+1)$-dimensional spacetime with Minkowski metric $\eta^{\mu\nu}=\left(+,-,-\right)$.

We start by considering the Lagrangian density for the Lee-Wick electrodynamics \cite{DiracString4,LeeWick},
\begin{eqnarray}
\label{LWl}
{\cal L}&=& -\frac{1}{4}F_{\mu\nu}F^{\mu\nu}-\frac{1}{2\xi}\left(\partial_{\mu}A^{\mu}\right)^{2}
-\frac{1}{4m^{2}}F_{\mu\nu}\partial_{\alpha}\partial^{\alpha} F^{\mu\nu}\nonumber\\
&
&-J^{\mu}A_{\mu} \ ,
\end{eqnarray}
where $A^{\mu}$ is the vector potential, $F^{\mu\nu}=\partial^{\mu}A^{\nu}-\partial^{\nu}A^{\mu}$ is the field strength, $J^{\mu}$ is the external source, $\xi$ is a gauge fixing parameter and $m$ is a parameter with mass dimension. 
The propagator in the Feynman gauge, $\xi=1$, for the theory in (\ref{LWl}) is given by \cite{DiracString4,LeeWick}
\begin{eqnarray}
\label{LWprop}
D^{\mu\nu}\left(x,y\right)&=&\int\frac{d^{3}p}{(2\pi)^{3}}\left(\frac{1}{p^{2}-m^{2}}-\frac{1}{p^{2}}\right)\nonumber\\
&
&\times\left(\eta^{\mu\nu}-\frac{p^{\mu}p^{\nu}}{m^{2}}\right)e^{-ip\cdot(x-y)} \ .
\end{eqnarray}

For a quadratic theory, like the one in (\ref{LWl}), the interaction between stationary field sources can be obtained from expression \cite{Zee,BaroneHidalgo1,BaroneHidalgo2,Camilo,Medeiros,DiracString1}   
\begin{equation}
\label{energyS}
E=\frac{1}{2T}\int\int d^{3}x\ d^{3}y J^{\mu}(x)D_{\mu\nu}(x,y)J^{\nu}(y)\ ,
\end{equation}
where $T$ is the time variable and it is implicit the limit $T\rightarrow\infty$.

First let us consider the interaction between two point-like field sources in two dimensions, which is described by the following external source
\begin{eqnarray}
\label{corre1Em}
J^{CC}_{\mu}({\bf x})=\lambda_{1}\eta_{\ \mu}^{0}\delta^{2}\left({\bf x}-{\bf a}_ {1}\right)+\lambda_{2}\eta_{\ \mu}^{0}\delta^{2}\left({\bf x}-{\bf a}_ {2}\right) \ ,
\end{eqnarray}
where the position of the charges are given by the spatial vectors ${\bf a}_ {1}$ and ${\bf a}_ {2}$ and the super-index $CC$ means that we have a system composed by two point-like charges in the planar Lee-Wick electrodynamics.

From now on in this paper, for simplicity, we shall refer to stationary point-like sources in $(2+1)$ dimensions, like the ones in (\ref{corre1Em}), as charges. Besides, we shall use the symbol $\lambda$ to designate the intensity of this kind of source.

Substituting (\ref{LWprop}) and (\ref{corre1Em}) in (\ref{energyS}), discarding the self-interacting contributions, we obtain
\begin{eqnarray}
\label{eeeenergy}
E^{CC}=\lambda_{1}\lambda_{2}\left(\int\frac{d^{2}{\bf p}}{\left(2\pi\right)^{2}}\frac{e^{i{\bf p}\cdot{\bf a}}}{{\bf p}^{2}}-\int\frac{d^{2}{\bf p}}{\left(2\pi\right)^{2}}\frac{e^{i{\bf p}\cdot{\bf a}}}{{\bf p}^{2}+m^{2}}\right) \ ,
\end{eqnarray}
with ${\bf{a}}={\bf {a}}_{1}-{\bf {a}}_{2}$ standing for the distance between the two charges.

Performing the relevant integrals in (\ref{eeeenergy}), we arrive at
\begin{equation}
\label{eeeenergy2}
E^{CC}=-\frac{\lambda_{1}\lambda_{2}}{2\pi}\left[\ln\left(\frac{a}{a_{0}}\right)+K_{0}\left(ma\right)\right]\ ,
\end{equation}
where $a=\mid\bf{a}\mid$, $a_{0}$ is an arbitrary constant length scale and $K_{0}\left(ma\right)$ stands for the K-Bessel function \cite{Arfken}. This result is already known in the literature \cite{DiracString4} and we present it in this letter just for completeness and to point out some peculiarities in comparison with the $(3+1)$-dimensional case. 

The interaction force between the charges reads
\begin{equation}
\label{FCC3d}
F^{CC}=-\frac{dE^{CC}}{da}= \frac{\lambda_{1}\lambda_{2}}{2\pi a}\left[1-\left(ma\right)K_{1}\left(ma\right)\right] \ .
\end{equation}

The first term between brackets on the right hand side of the Eq. (\ref{FCC3d}) is the usual Coulomb interaction obtained in $3d$ Maxwell electrodynamics. The second one is a contribution due to the parameter $m$, which falls down when $m$ or $a$ increases. We notice that the force is repulsive for charges with the same signal and attractive otherwise.  

\begin{figure}[!h]
\centering \includegraphics[scale=0.37]{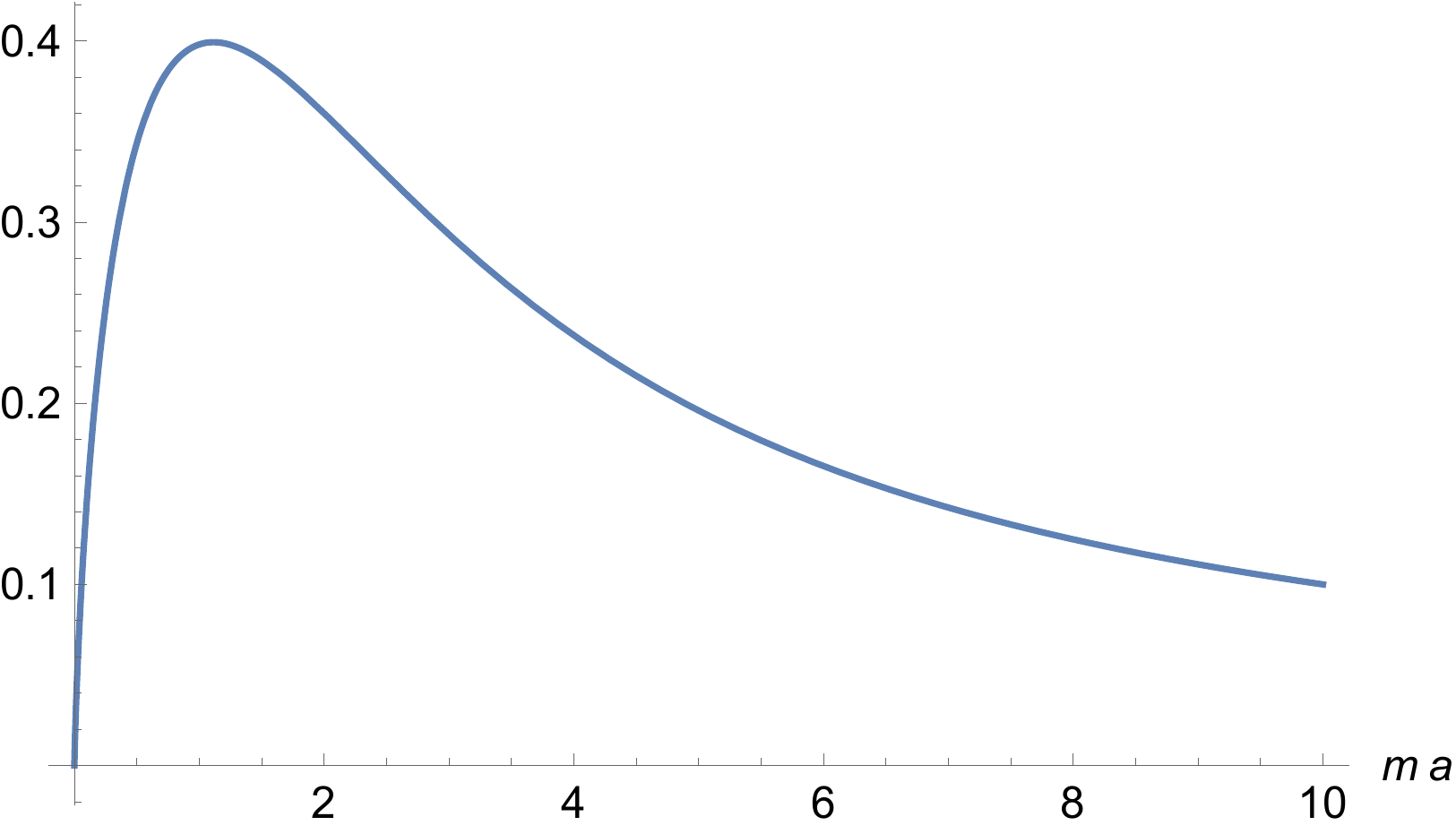} \caption{Plot for $\frac{2\pi F^{CC}}{m\lambda_{1}\lambda_{2}}$ in Eq. (\ref{FCC3d}), as a function of $ma$.}
\label{CC3d} 
\end{figure}

In Fig. \ref{CC3d}, we have a plot for the force (\ref{FCC3d}) multiplied by $\frac{2\pi}{m\lambda_{1}\lambda_{2}}$. We highlight that this force has a global maximum around $ma\cong 1,14$, and vanishes when $a\rightarrow0$ .

At this point we highlight some differences in the behavior of the forces between point-like sources the in Lee-Wick electrodynamics when we compare the $(3+1)$-dimensional case with the planar situation.

The interaction force between two point-like charges $q_{1}$ and $q_{2}$ placed at a distance $a$ apart in the $(3+1)$-dimensional Lee-Wick electrodynamics can be computed by taking the derivative (with an overall minus sign) of Eq. (16) of Ref. \cite{DiracString4} with respect to distance $a$, 
\begin{eqnarray}
\label{FCC4d}
F^{CC}_{(3+1)}=\frac{q_{1}q_{2}}{4\pi a^{2}}\left[1-e^{-ma}-\left(ma\right)e^{-ma}\right] \ , 
\end{eqnarray}

\begin{figure}[!h]
\centering \includegraphics[scale=0.37]{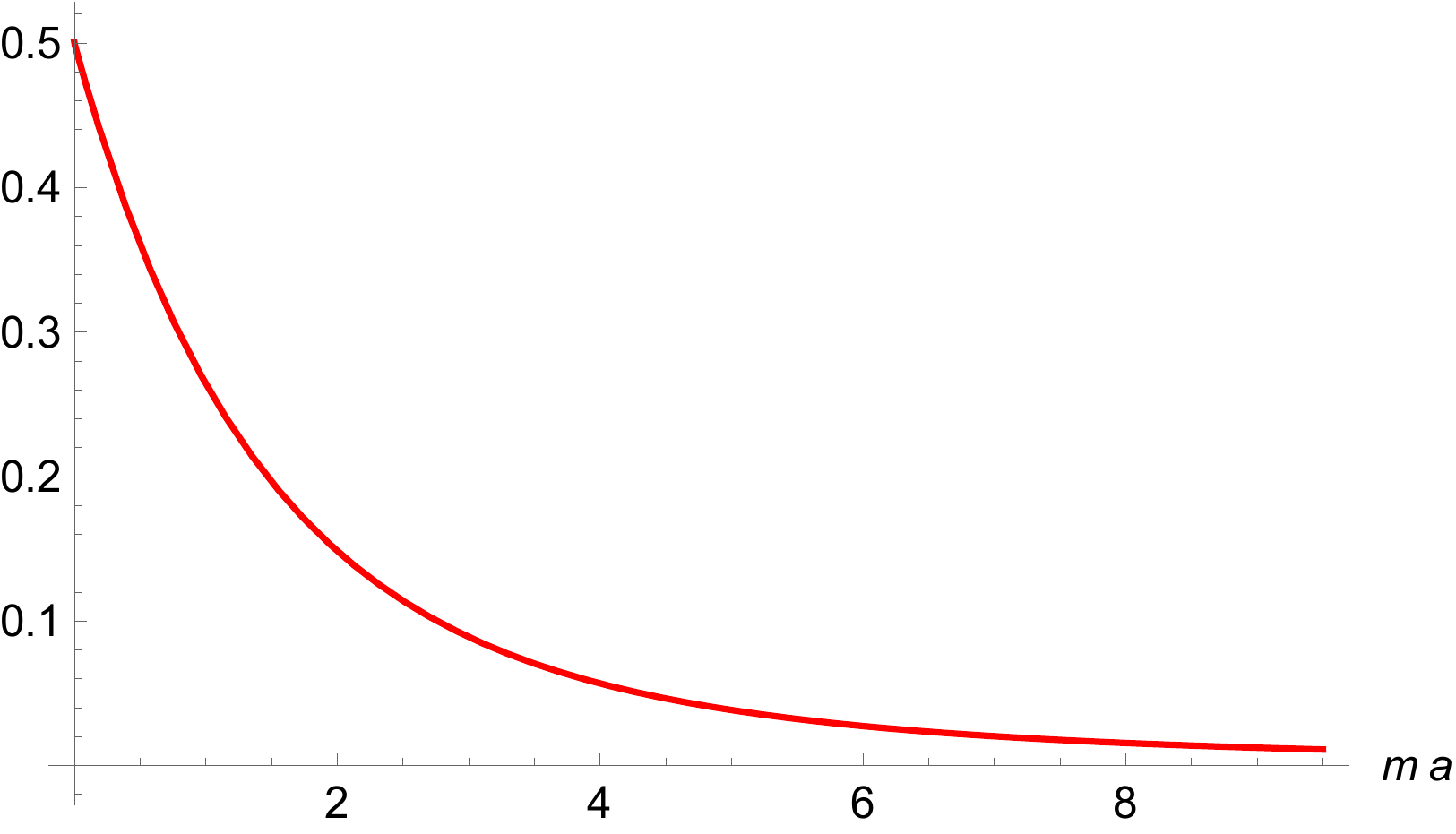} \caption{Plot for $\frac{4\pi F^{CC}_{(3+1)}}{m^{2}q_{1}q_{2}}$ in Eq. (\ref{FCC4d}), as a function of $ma$.}
\label{CC4d} 
\end{figure}

In Fig. \ref{CC4d}, we have a plot for the force (\ref{FCC4d}) multiplied by $\frac{4\pi}{m^{2}q_{1}q_{2}}$, where we can see that when $a\rightarrow0$, this force is finite and equal to $\frac{m^{2}q_{1}q_{2}}{8\pi}$. The curve in the graphic (\ref{CC4d}) goes to $1/2$ when $ma=0$.

Comparing the figures \ref{CC3d} and \ref{CC4d}, we notice that the behavior of the interaction force between two point-like charges in $3d$ Lee Wick electrodynamics is notably different from the correspondent one obtained in $4d$ dimensions. In the $4d$ case, the force is monotonic, falls down when $ma$ increases and exhibits a nonzero value for $ma=0$. For the planar setup, the force exhibits a global maximum around $ma\cong1.44$, vanishes for $ma=0$ and goes to zero when $ma\to\infty$. In both cases the force is always repulsive for charges with the same signal and attractive otherwise.  

It is well-known that the presence of a perfectly conducting line imposes a boundary condition on the gauge field in such a way that the Lorentz force on the line must vanish, as discussed in Ref. \cite{LeeWick}.

Now, let us consider the presence of a perfectly conducting line in $(2+1)$-dimensional Lee-Wick electrodynamics. We take a coordinate system where the conductor lies on the line $x^{2}=a$. In this case, the condition that makes the Lorentz force to vanish along the conducting line is achieved by
\begin{equation}
\label{condition11}
n^{\mu  \ *}F_{\mu}|_{x^{2}=a}=0 \Rightarrow \epsilon_{2}^{\ \nu\lambda}\partial_{\nu}A_{\lambda}\left(x\right)|_{x^{2}=a}=0 \ ,
\end{equation}
where $n^{\mu}=\eta_{2}^{\ \mu}=\left(0,0,1\right)$ is the Lorentz three-vector normal to the conducting line and $ ^{*}F^{\mu}=(1/2)\epsilon^{\mu\nu\lambda}F_{\nu\lambda}$ is the dual field strength, with $\epsilon^{\mu\nu\alpha}$ standing for the Levi-Civita tensor $(\epsilon^{012}=1)$.
 
By using the functional formalism employed in \cite{LHCHLOFAB,Bordag,LeeWick,LHCFABplate,FABplate}, we can write the functional generator as follows
\begin{eqnarray}
\label{fgen1}
Z_{C}\left[J\right]=\int {\cal{D}}A_{C} \ e^{i\int d^{3}x \ \cal{L}} \ ,
\end{eqnarray}
where the sub-index $C$ means that we are integrating out in all field configurations which satisfy the condition (\ref{condition11}). This restriction is achieved by introducing a delta functional, which is not equal to zero only where the restrictions (\ref{condition11}) are satisfied, as follows
\begin{eqnarray}
\label{fgen2}
Z_{C}\left[J\right]=\int {\cal{D}}A \ \delta\left[{^{*}F}_{2}\left(x\right)|_{x^{2}=a}\right] \ e^{i\int d^{3}x \ \cal{L}} \ .
\end{eqnarray}

Now we use the Fourier representation for the delta functional
\begin{eqnarray}
\label{fgen3}
\delta\left[{^{*}F}_{2}\left(x\right)|_{x^{2}=a}\right]&=&\int {\cal{D}}B\exp\Bigl[i
\int d^{3}x\ \delta\left(x^{2}-a\right)\nonumber\\
&
&\times B\left(x_{\parallel}\right) {^{*}F_{2}\left(x\right)}\Bigr] \ ,
\end{eqnarray}
where $B\left(x_{\parallel}\right)$ is an auxiliary scalar field and $x_{\parallel}^{\mu}=\left(x^{0},x^{1},0\right)$ means that we have only the coordinates parallel to the conducting line. (In the (3+1) dimensional case, we have a similar expression, but with an auxiliary vector field \cite{LeeWick}). 

Performing similar steps that was employed in Ref. \cite{LHCHLOFAB}, we can write the functional generator in the following way
\begin{eqnarray}
\label{fgen5}
Z_{C}\left[J\right]=Z\left[J\right]{\bar{Z}}\left[J\right] \ ,
\end{eqnarray}
where $Z\left[J\right]$ is the usual functional generator for the gauge field,
\begin{eqnarray}
\label{fgen6}
Z\left[J\right]&=&Z\left[0\right]\exp\Bigl[-\frac{i}{2}\int d^{3}x \ d^{3}y \ J^{\mu}\left(x\right)D_{\mu\nu}
\left(x,y\right)\nonumber\\
&
&\times J^{\nu}\left(y\right)\Bigr] \ ,
\end{eqnarray}
and ${\bar{Z}}\left[J\right]$ is a contribution due to the scalar field $B\left(x_{\parallel}\right)$,  
\begin{eqnarray}
\label{fgen7}
{\bar{Z}}\left[J\right]&=&\int{\cal{D}}B\exp\left[i\int d^{3}x \ \delta
\left(x^{2}-a\right)I\left(x\right)B\left(x_{\parallel}\right)\right] \nonumber\\
&
&\times\exp\Bigl[-\frac{i}{2}\int d^{3}x \ d^{3}y \ \delta\left(x^{2}-a\right)
\delta\left(y^{2}-a\right)\nonumber\\
&
&\times B\left(x_{\parallel}\right)W\left(x,y\right)
B\left(y_{\parallel}\right)\Bigr] \ ,
\end{eqnarray}
where we defined
\begin{eqnarray}
\label{defi1}
I\left(x\right)&=&-\int d^{3}y \ \epsilon_{2}^{\ \gamma\alpha}
\left(\frac{\partial}{\partial x^{\gamma}}D_{\alpha\mu}\left(x,y\right) 
\right)J^{\mu}\left(y\right) \ , \nonumber\\   
W\left(x,y\right)&=&\epsilon_{2}^{\ \gamma\alpha}
\epsilon_{2}^{\ \beta\lambda}\frac{\partial^{2}D_{\lambda\alpha}\left(x,y\right)}
{\partial x^{\beta}\partial y^{\gamma}} \ .
\end{eqnarray}

Substituting (\ref{defi1}) and (\ref{LWprop}) into (\ref{fgen7}), using the fact that \cite{LHCHLOFAB,LeeWick,LHCFABplate} 
\begin{eqnarray}
\label{int}
\int \frac{dp^{2}}{2\pi}\frac{e^{i p^{2}\left(x^{2}-y^{2}\right)}}
{p^{\mu}p_{\mu}-m^{2}}&=&-\frac{i}{2\Gamma} \ e^{i\Gamma\mid x^{2}-y^{2}\mid} \ , \nonumber\\
\int \frac{dp^{2}}{2\pi}\frac{e^{i p^{2}\left(x^{2}-y^{2}\right)}}
{p^{\mu}p_{\mu}}&=&-\frac{i}{2L} \ e^{iL\mid x^{2}-y^{2}\mid} \ ,
\end{eqnarray}
where $p^{2}$ stands for the momentum component perpendicular to the conducing line, $\Gamma=\sqrt{p_{\parallel}^{2}-m^{2}}$ and $L=\sqrt{p_{\parallel}^{2}}$, defining  the parallel momentum to the line $p_{\parallel}^{\mu}=\left(p^{0},p^{1},0\right)$ and the parallel metric 
\begin{eqnarray}
\label{etap}
\eta_{\parallel}^{\mu\nu}=\eta^{\mu\nu}-\eta_{\ 2}^{\mu}\eta^{\nu 2} \ ,
\end{eqnarray}
one can write the expression (\ref{fgen7}) as follows
\begin{eqnarray}
\label{fgen8}
{\bar{Z}}\left[J\right]&=&{\bar{Z}}\left[0\right]\exp\Bigl[-\frac{i}{2}\int d^{3}x 
\ d^{3}y \ J^{\mu}\left(x\right){\bar{D}}_{\mu\nu}\left(x,y\right)\nonumber\\
&
&\times J^{\nu}\left(y\right)\Bigr] \ ,
\end{eqnarray}
where we defined the function
\begin{eqnarray}
\label{LWprpline}
&&{\bar{D}}^{\mu\nu}\left(x,y\right)=-\frac{i}{2}\int \frac{d^{2}p_{\parallel}}
{\left(2\pi\right)^{2}} \left(\eta_{\parallel}^{\mu\nu}-\frac{p_{\parallel}^{\mu}p_{\parallel}^{\nu}}{p_{\parallel}^{2}}\right)\frac{1}{\left(\frac{1}{L}-\frac{1}{\Gamma}\right)}\nonumber\\
&
&\times\exp\left[{-i p_{\parallel}\cdot\left(x_{\parallel}
-y_{\parallel}\right)}\right]
\left(\frac{e^{iL\mid x^{2}-a\mid}}{L}-\frac{e^{i\Gamma\mid x^{2}-a\mid}}{\Gamma}\right)\nonumber\\
&
&\times\left(\frac{e^{iL\mid y^{2}-a\mid}}{L}-\frac{e^{i\Gamma\mid y^{2}-a\mid}}{\Gamma}\right) \ .
\end{eqnarray}

Substituting (\ref{fgen8}) and (\ref{fgen6}) in (\ref{fgen5}), the functional generator for the $(2+1)$-dimensional Lee-Wick electrodynamics in the presence of a perfectly conducting line reads
\begin{eqnarray}
\label{fgen9}
Z_{C}\left[J\right]&=&Z_{C}\left[0\right]\exp\Bigl[-\frac{i}{2}
\int d^{3}x \ d^{3}y \ J^{\mu}\left(x\right)\Bigl(D_{\mu\nu}
\left(x,y\right)\nonumber\\
&
&+{\bar{D}}_{\mu\nu}\left(x,y\right)\Bigr)J^{\nu}
\left(y\right)\Bigr] \ .
\end{eqnarray}

From the Eq. (\ref{fgen9}), one can identify the propagator of the $3d$ Lee-Wick electrodynamics in the presence of a conducting line as follows
\begin{eqnarray}
\label{prop3}
D_{C}^{\mu\nu}=D^{\mu\nu}\left(x,y\right)+{\bar{D}}^{\mu\nu}\left(x,y\right) \ .
\end{eqnarray}

The propagator (\ref{prop3}) is composed of the sum of the standard Lee-Wick propagator (\ref{LWprop}) with the correction (\ref{LWprpline}), which accounts for the presence of the perfectly conducting line. In the limit $m\rightarrow\infty$, the propagator
(\ref{LWprpline}) reduces to the same one as that we would have obtained with the planar Maxwell electrodynamics in the presence of a conducting line. It can be showed that the conducting line condition (\ref{condition11}) is satisfied by the propagator (\ref{prop3}), namely, $\epsilon_{2}^{\ \nu\lambda}\partial_{\nu}D_{C}^{\mu\nu}|_{x^{2}=a}=0$.

Now, let us investigate the interaction between a point-like charge and the conducting line. The interaction energy between a stationary field source and a conducting surface reads \cite{LHCHLOFAB,LeeWick,LHCFABplate,FAFEB,FAFEB2,Helder}
\begin{eqnarray}
\label{energy}
{{E}}=\frac{1}{2T}\int d^{3}x \ d^{3}y \ J^{\mu}\left(x\right)
{\bar{D}}_{\mu\nu}\left(x,y\right)J^{\nu}\left(y\right) \ .
\end{eqnarray}

Without loss of generality, we consider a charge located at position ${\bf{b}}=\left(0,b\right)$ whose external source is given by
\begin{eqnarray}
\label{source1}
J^{C}_{\mu}\left(x\right)=\lambda\eta^{0}_{\ \mu}\delta^{2}\left({\bf x}-{\bf b}\right) \ .
\end{eqnarray}
 
Substituting (\ref{source1}) and (\ref{LWprpline}) in (\ref{energy}), and then carrying out some straightforward manipulations, we arrive at
\begin{eqnarray}
\label{enerppar}
E^{LC}&=&-\frac{\lambda^{2}}{4\pi}\int^{\infty}_{0}d|{\bf{p}}_{\parallel}|\frac{\sqrt{{\bf{p}}_{\parallel}^{2}}\sqrt{{\bf{p}}_{\parallel}^{2}+m^{2}}}{\sqrt{{\bf{p}}_{\parallel}^{2}+m^{2}}-\sqrt{{\bf{p}}_{\parallel}^{2}}}\nonumber\\
&
&\times\left(\frac{e^{-R\sqrt{{\bf{p}}_{\parallel}^{2}}}}{\sqrt{{\bf{p}}_{\parallel}^{2}}}-\frac{e^{-R\sqrt{{\bf{p}}_{\parallel}^{2}+m^{2}}}}{\sqrt{{\bf{p}}_{\parallel}^{2}+m^{2}}}\right)^{2} \ ,
\end{eqnarray}
where super-index $LC$ means that we have a system consisting of a charge and a conducting line, and $R=\mid b-a\mid$ is the distance between the line and the charge. 

Performing the change of integration variable $p =|{\bf{p}}_{\parallel}|/m$, the expression (\ref{enerppar}) can be simplified as follows
\begin{eqnarray}
\label{cplatehoc}
E^{LC}&=&-\frac{\lambda^{2}}{4\pi}\int_{0}^{\infty} dp \ p\left[\left(p^{2}+1\right)+p\sqrt{p^{2}+1}\right]\nonumber\\
&
&\times\Biggl(\frac{e^{-2pmR}}{p^{2}}-2\frac{e^{-\left(p+\sqrt{p^{2}+1}\right)mR}}{p\sqrt{p^{2}+1}}\nonumber\\
&
&+\frac{e^{-2mR\sqrt{p^{2}+1}}}{p^{2}+1}\Biggr) \ .
\end{eqnarray}

Carrying out the integrals, we obtain that
\begin{eqnarray}
\label{contribution1}
&&\int_{0}^{\infty} dp \left[\left(p^{2}+1\right)+p\sqrt{p^{2}+1}\right]\frac{e^{-2pmR}}{p}\nonumber\\
&&\to-\ln\left(\frac{R}{R_{0}}\right)+\frac{1}{4\left(mR\right)^{2}}+\frac{\pi}{4mR}\Bigl[SH_{1}\left(2mR\right)\nonumber\\
&
&-Y_{1}\left(2mR\right)\Bigr] \ , \nonumber
\end{eqnarray}
\begin{eqnarray}
&&\int_{0}^{\infty} dp\ p\left[1+p\left(p^{2}+1\right)^{-1/2}\right]e^{-2mR\sqrt{p^{2}+1}}
\nonumber\\
&&=\frac{K_{1}\left(2mR\right)}{2mR}+\frac{e^{-2mR}}{4\left(mR\right)^{2}}\left(1+2mR\right) \ , \nonumber
\end{eqnarray}
\begin{eqnarray}
&&-2\int_{0}^{\infty} dp\left(p+\sqrt{p^{2}+1}\right)e^{-\left(p+\sqrt{p^{2}+1}\right)mR}\nonumber\\
&&=-e^{-mR}\left(\frac{1}{\left(mR\right)}+\frac{1}{\left(mR\right)^{2}}\right)-Ei\left(1,mR\right) \ , 
\end{eqnarray}
where $R_{0}$ is an arbitrary finite constant with dimension of length and $Y$, $SH$, $K$, $Ei$ stand for the Bessel function of second kind, the Struve function, $K$-Bessel function and exponential integral function, respectively \cite{Arfken}. In the first integral (\ref{contribution1}) we have added a $R$-independent term and discarded some contributions which do not depend on $R$.

Finally, substituting (\ref{contribution1})  in (\ref{cplatehoc}), we arrive at
\begin{eqnarray}
\label{cplate2}
E^{LC}&=&-\frac{\lambda^{2}}{4\pi}\Biggl[-\ln\left(\frac{R}{R_{0}}\right)+\frac{1}{4\left(mR\right)^{2}}\nonumber\\
&
&+\frac{\pi}{4mR}\Bigl[SH_{1}\left(2mR\right)
-Y_{1}\left(2mR\right)\Bigr]\nonumber\\
&
&-e^{-mR}\left(\frac{1}{\left(mR\right)}+\frac{1}{\left(mR\right)^{2}}\right)-Ei\left(1,mR\right)\nonumber\\
&
&+\frac{K_{1}\left(2mR\right)}{2mR}+\frac{e^{-2mR}}{4\left(mR\right)^{2}}\left(1+2mR\right)\Biggr] \ .
\end{eqnarray}

Eq. (\ref{cplate2}) is an exact result and gives the interaction energy between a point-like charge and the conducting line for the $3d$ Lee-Wick electrodynamics. The first term on the right hand side is the charge-line interaction obtained in the standard planar Maxwell theory. The remaining contributions are corrections due to the parameter $m$ and fall down when $R$ increases faster that the first term.

The interaction force between the conducting line and the charge can be computed from the Eq. (\ref{cplate2}), resulting in
\begin{eqnarray}
\label{FFCP}
F^{LC}&=&-\frac{\lambda^{2}}{4\pi R}\Biggl[1+\frac{1}{2\left(mR\right)^{2}}\nonumber\\
&
&-\frac{\pi}{2}\left(Y_{2}\left(2mR\right)+SH_{0}\left(2mR\right)-\frac{SH_{1}\left(2mR\right)}{\left(mR\right)}\right)\nonumber\\
&
&+K_{2}\left(2mR\right)
-2e^{-mR}\left(1+\frac{1}{\left(mR\right)}+\frac{1}{\left(mR\right)^{2}}\right)\nonumber\\
&
&+e^{-2mR}\left(1+\frac{1}{\left(mR\right)}+\frac{1}{2\left(mR\right)^{2}}\right)\Biggr] \ .
\end{eqnarray}

\begin{figure}[!h]
\centering \includegraphics[scale=0.37]{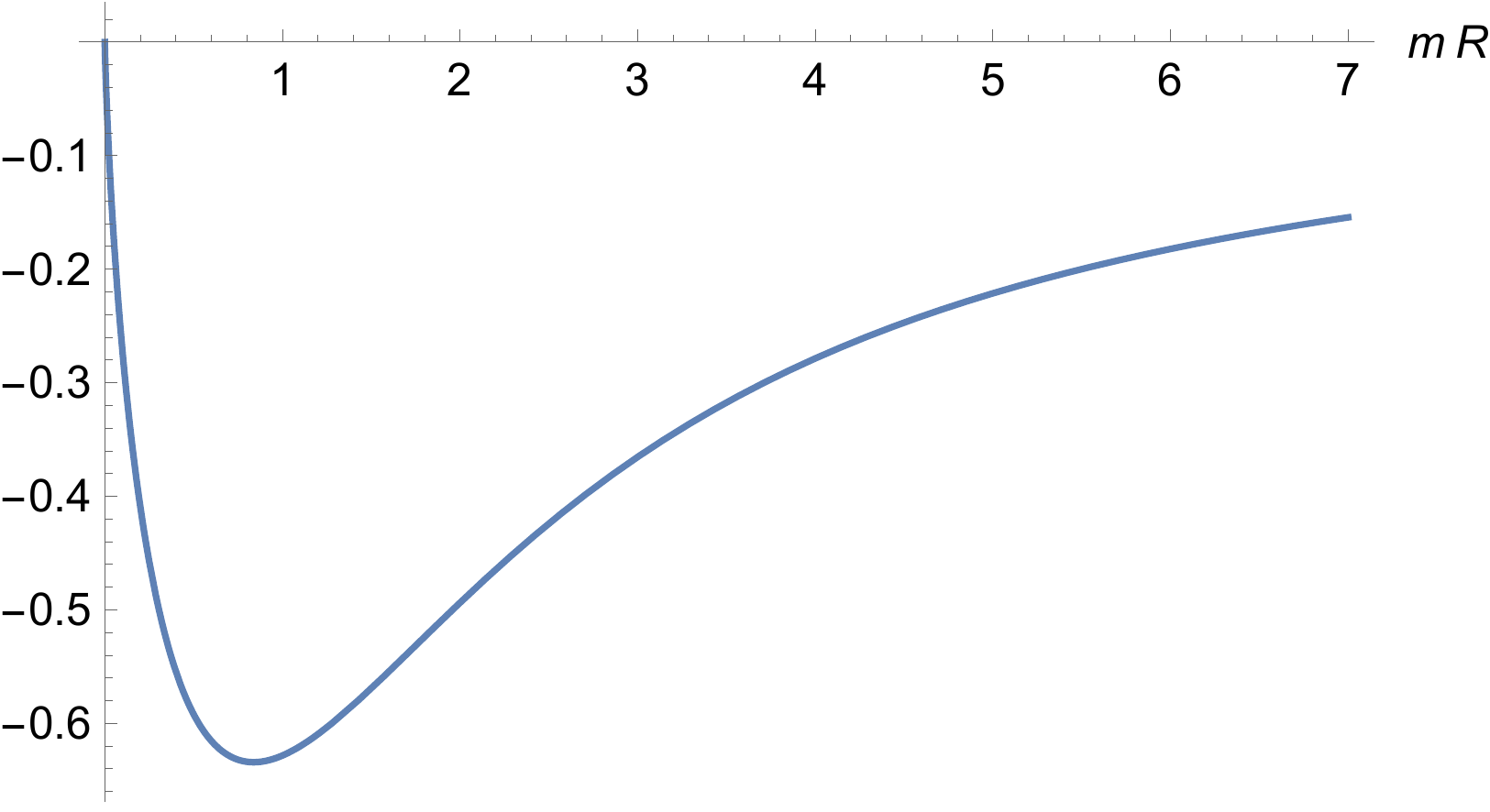} \caption{Plot for $\frac{4\pi F^{LC}}{m\lambda^{2}}$ in Eq. (\ref{FFCP}), as a function of $mR$.}
\label{LinhaCarga} 
\end{figure}

The first term on the right hand side of the expression (\ref{FFCP}) is the interaction force between the charge $\lambda$ and its image $-\lambda$, placed at a distance $2R$ apart, obtained in $(2+1)$-dimensional Maxwell electrodynamics. The $m$-dependent contribution falls down when $mR$ increases. In Eq. (\ref{FFCP}) the term between brackets is positive, what implies that the interaction force is always attractive. In Fig. (\ref{LinhaCarga}) we have a plot for the force (\ref{FFCP}) multiplied by $\frac{4\pi}{m\lambda^{2}}$. We can see that the interaction force has a global minimum around $mR\cong 0.82$. 

It is important to highlight that the results (\ref{cplate2}) and (\ref{FFCP}) are valid just for $mR\neq 0$. The case where $R=0$ must be treated carefully. Taking the derivative of Eq. (\ref{cplatehoc}) with respect to $R$ and evaluating for $R=0$, we have zero as result, what implies that the interaction force between the conducting line and the point-like charge vanishes when the charge is placed on the line. Now taking the limit $R\rightarrow0$ in Eq. (\ref{FFCP}) we also obtain zero as result. Therefore, the interaction force is finite and continuous at $R=0$, this fact is evinced in Fig. (\ref{LinhaCarga}).  

The interaction force between two point-like charges for the $3d$ Lee-Wick electrodynamics (theory without the conducting line) is given by (\ref{FCC3d}). For the special case where we have two opposite charges, $\lambda_{1}=\lambda$, $\lambda_{2}=-\lambda$, placed at a distance $2R$ apart, we obtain
\begin{equation}
\label{forLWCC}
F^{CC}=-\frac{\lambda^{2}}{4\pi R}\left[1-\left(2mR\right)K_{1}\left(2mR\right)\right] \ ,
\end{equation}
where the super-index $CC$ means that we have the interaction between two point-like charges.

We can verify that the force (\ref{forLWCC}) is very different from (\ref{FFCP}). So, we notice that the image method is not valid for the $3d$ Lee-Wick electrodynamics for the conducting line condition (\ref{condition11}). An opposite situation occurs in Maxwell-Chern-Simons electrodynamics, where the image method is valid for point-like charges \cite{LHCHLOFAB}.

\begin{figure}[!h]
\centering \includegraphics[scale=0.37]{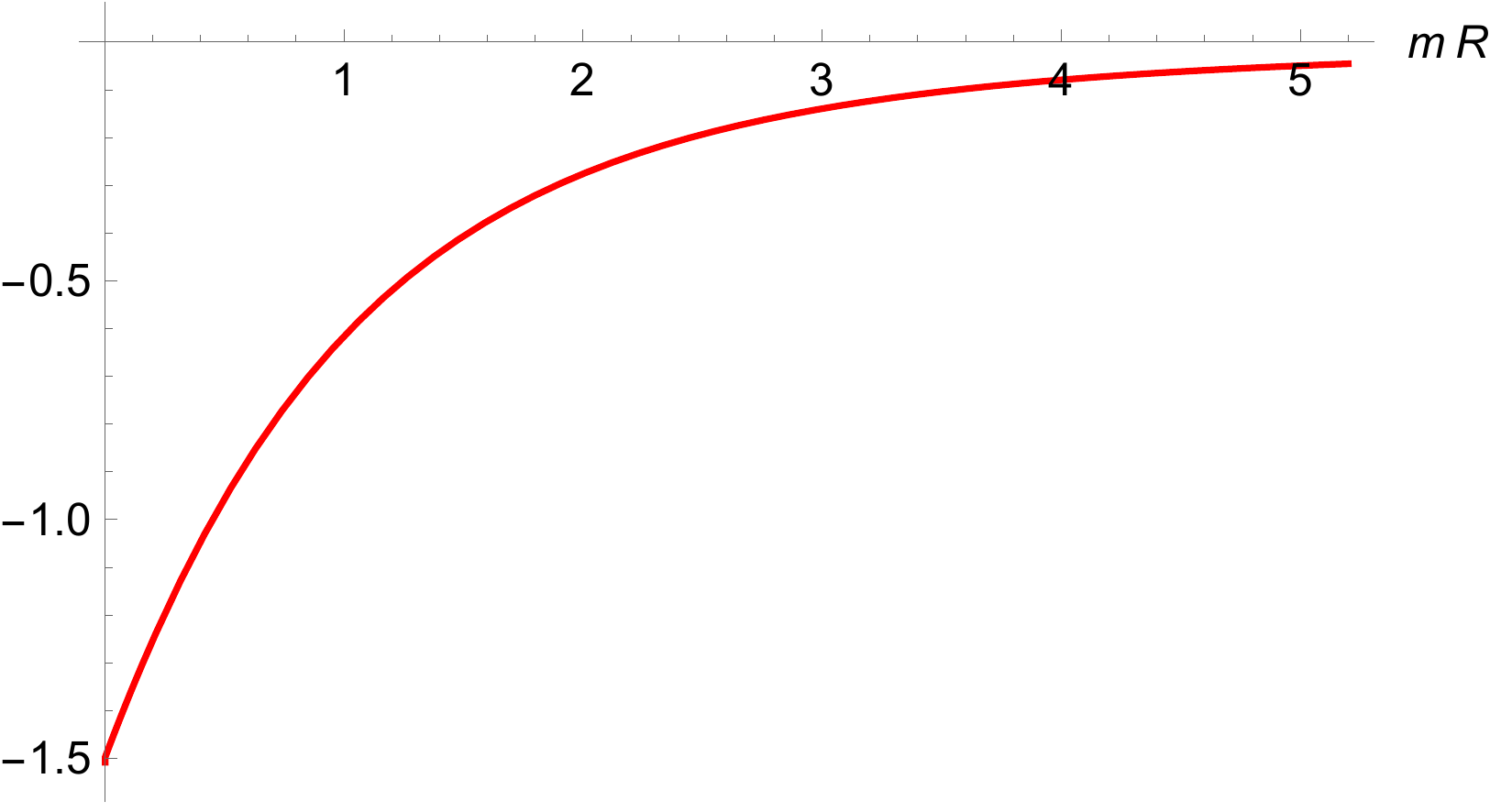} \caption{Plot for $\frac{16\pi F^{PC}}{m^{2}q^{2}}$ in Eq. (37) of Ref. \cite{LeeWick}, in $(3+1)$ dimensions, as a function of $mR$.}
\label{plate4d} 
\end{figure}

In Ref. \cite{LeeWick} the Lee-Wick electrodynamics in $3+1$ dimensions was investigated in the presence of a perfectly conducting plate, where it was shown that the image method is not valid for point-like charges. Using the expression (37) of reference \cite{LeeWick} we can obtain the interaction force between the plate and the charge, $F_{PC}$, whose behavior is evinced in Fig. (\ref{plate4d}) where we have a plot for $\frac{16\pi F^{PC}}{m^{2}q^{2}}$, where $q$ is the charge intensity . 
  
Comparing the figures (\ref{LinhaCarga}) and (\ref{plate4d}), we notice that the behavior of the interaction force between the point-like charge and a conducting line for the the $3d$ Lee-Wick electrodynamics is different from the one obtained in $4d$ Lee-Wick electrodynamics, where the analogue of a conducting line is just a conducting plate. In Fig. (\ref{plate4d}) we can see that $\lim_{R\rightarrow0}F^{PC}=-3m^{2}q^{2}/(32\pi)$. However, in $R=0$, we have $F^{PC}=0$, this fact can be verified by taking the derivative of Eq. (30) of Ref. \cite{LeeWick} with respect to $R$ and then, taking $R=0$. So, we notice that $F^{PC}$ is discontinuous when the charge is placed on the plate, what is an opposite situation to the one found in $(2+1)$ dimensional Lee-Wick electrodynamics.

It is important to mention that the results obtained throughout this letter cannot be directly predicted from the ones of Ref. \cite{LeeWick} before the calculations are performed.

In summary, in the present letter we have investigated some aspects of the $(2+1)$-dimensional Lee-Wick electrodynamics in the vicinity of a perfectly conducting line. Specifically, the propagator for the Lee-Wick field and the interaction force between the conducting line and a point-like charge were computed. We have showed that the behavior of the interaction forces are different from that ones found in $(3+1)$-dimensional Lee-Wick electrodynamics. We have also verified that the image method remains not valid in $3d$ Lee-Wick electrodynamics.

{\bf Acknowledgments:}\\
For financial support, L.H.C. Borges thank to CAPES (Brazilian agency) and  F.A. Barone thanks to CNPq (Brazilian agency) under the grant P313978/2018-2 .

\end{document}